\documentclass[11pt]{article}   
   
\usepackage{color}
\definecolor{indigo}{RGB}{0,0,120}
\usepackage[colorlinks=true, linkcolor=indigo, citecolor=blue, urlcolor=blue]{hyperref}

\def\tr{\;{\rm tr}\;}

\def\fl{\noindent}

\newcommand{\beq}{\begin{equation}}
\newcommand{\eeq}{\end{equation}}
\newcommand{\beqs}{\begin{eqnarray}}
\newcommand{\eeqs}{\end{eqnarray}}

\newcommand{\ov}[1]{\frac{1}{#1}}
\newcommand{\fr}[2]{\frac{#1}{#2}}

			\def\eps{\epsilon} 
  
\def\la{\lambda}		\def\sig{\sigma}

\def\vf{\varphi}		
\def\tht{\theta}	
		\def\om{\omega}

\usepackage{amsmath,amssymb} 

\usepackage{graphicx}

\usepackage{calligra} 
\DeclareMathAlphabet{\mathcalligra}{T1}{calligra}{m}{n}
\DeclareFontShape{T1}{calligra}{m}{n}{<->s*[2.2]callig15}{}
\newcommand{\scripty}[1]{\ensuremath{\mathcalligra{#1}}}

\newcount\colveccount  
\newcommand*\colvec[1]{\global\colveccount#1  \begin{pmatrix} \colvecnext} \def\colvecnext#1{#1 \global\advance\colveccount-1
        \ifnum\colveccount>0 \\ \expandafter\colvecnext
        \else \end{pmatrix} \fi}

\newcommand{\err}{\scripty{r}}

\usepackage{subcaption}

\begin{document}



\title{\normalsize 
\hfill {\tt arXiv:1810.01317} \\
\vskip 0.1mm \LARGE
Stability and chaos in the classical three rotor problem
\\
}
\author{{\sc Govind S. Krishnaswami and Himalaya Senapati}
\\ \\ \small
 Chennai Mathematical Institute,  SIPCOT IT Park, Siruseri 603103, India
\\ \small
 Email: {\tt govind@cmi.ac.in, himalay@cmi.ac.in}}

\date{September 24, 2019 \\ \href{https://www.ias.ac.in/article/fulltext/conf/002/0020}{Published} in Indian Academy of Sciences Conference Series {\bf 2}(1), 139 (2019)}

\maketitle

We study the equal-mass classical three rotor problem, a variant of the three body problem of celestial mechanics. The quantum $N$-rotor problem has been used to model chains of coupled Josephson junctions and also arises via a partial continuum limit of the Wick-rotated XY model. In units of the coupling, the energy serves as a control parameter. We find periodic `pendulum' and `breather' orbits at all energies and choreographies at relatively low energies. They furnish analogs of the Euler-Lagrange and figure-8 solutions of the planar three body problem. Integrability at very low energies gives way to a rather marked transition to chaos at $E_c \approx 4$, followed by a gradual return to regularity as $E \to \infty$. We find four signatures of this transition: (a) the fraction of the area of Poincar\'e surfaces occupied by chaotic sections rises sharply at $E_c$, (b) discrete symmetries are spontaneously broken at $E_c$, (c) $E=4$ is an accumulation point of stable to unstable transitions in pendulum solutions and (d) the Jacobi-Maupertuis curvature goes from being positive to having both signs above $E=4$. Moreover, Poincar\'e plots also reveal a regime of global chaos slightly above $E_c$.

{{\bf Keywords}: three-body problem, periodic orbits, stability, accumulation of phase transitions, transition to chaos, global chaos, choreographies.}

\footnotesize
\tableofcontents
\normalsize

\section{Introduction}

The classical 3 body problem arose in an attempt to understand the effect of the Sun on the Moon's Keplerian orbit around the Earth. It led to the discovery of chaos in the hands of Poincar\'e and continues to be a fertile area of research \cite{gutzwiller-three-body, chenciner-montgomery, gskhs-3body}. We study a variant, the classical 3 rotor problem, where 3 particles with equal masses move on a circle subject to pairwise attractive cosine potentials. While the case of two rotors reduces to that of a pendulum, the quantum $N$-rotor problem may be obtained via a Wick rotation of an anisotropic continuum limit of the XY lattice model of statistical mechanics \cite{sachdev, gskhs-3rotor} and has been used in the physics of coupled Josephson junctions \cite{sondhi}. Thus, in a suitable $N \to \infty$ limit, it is related to the sine-Gordon field \cite{sachdev}. On the other hand, the few rotor problems are largely unexplored, and as we will show by studying the case $N=3$, they display rich dynamics with novel signatures of chaos. Moreover, we will argue that collisional and non-collisional singularities familiar from gravitational $N$-body problems \cite{saari-xia} are absent in the 3 rotor problem allowing us to study periodic solutions, choreographies, instabilities and the transition to chaos in a simpler context.


\section{The classical three-rotor problem}

We study the problem of three coupled rotors interacting via cosine potentials defined by the Lagrangian
	\beq
	L_\text{tot} = \sum_{i=1}^3  \left[ \fr{ m r^2}{2} \dot\tht_i^2 - g [1 - \cos\left(\tht_i-\tht_{i+1} \right) ] \right] = K - V
	\eeq
where $\tht_{1,2,3}$ are the angular positions of the three rotors with $\tht_4 \equiv \tht_1$. We assume `ferromagnetic' coupling $g > 0$ where rotors attract each other. The phase space $M^6$ is the cotangent bundle of the three-torus configuration space with momenta $\pi_i = m r^2 \dot \tht_i$ and corresponding Hamiltonian $H_{\rm tot} = K + V$.
 Physical quantities may be non-dimensionalized using the constants $m$, $r$ and $g$. We will see that the energy in units of $g$ serves as a useful organizing parameter in discussing the dynamics. The Hamiltonian vector field defined by
	\beq
	\dot \tht_i = \pi_i/mr^2 
	\quad \text{and} \quad 
	\dot \pi_i = g \sin (\tht_{i-1}-\tht_i) - g \sin (\tht_i-\tht_{i+1})
	\label{e:hamilton-eom-theta}
	\eeq
is smooth everywhere on $M^6$ so that particles can pass through each other without singularities. Since the potential $V$ is non-negative, momenta are bounded and energy level sets are compact 5d sub-manifolds of $M^6$ without boundaries. Consequently, there are no `non-collisional' singularities where $\pi_i$ or $\tht_i$ diverge in finite time. This implies existence and uniqueness of solutions to (\ref{e:hamilton-eom-theta}) for all times.

It is convenient to define a centre of mass (CM) coordinate $\vf_0$ and relative coordinates $\vf_{1,2}$:
	\beq
	\vf_0 = (\tht_1 + \tht_2 + \tht_3)/3 	\quad \text{and} \quad 
	\vf_{1,2} = \tht_{1,2} - \tht_{2,3}.
	\eeq
The $2\pi$-periodicity of the $\tht$s implies that $\vf_0$ and $\vf_{1,2}$ are $2\pi$ and $6\pi$ periodic. However, we may take the fundamental region to be $[0, 2\pi]^3$ since $(\vf_0, \vf_1 - 2\pi, \vf_2)$, $(\vf_0, \vf_1, \vf_2 + 2\pi)$ and $(\vf_0 +2\pi/3, \vf_1, \vf_2)$ are identical configurations. Thus, the boundary conditions on this fundamental domain are not quite periodic. However, $\vf_1$ and $\vf_2$ evolve independently of $\vf_0$ so that they may be taken to be periodic coordinates on a 2-torus $[0,2\pi]^2$. On the other hand, when $\vf_1 \mapsto \vf_1 \pm 2\pi$ or $\vf_2 \mapsto \vf_2 \mp 2\pi$, the CM variable $\vf_0 \mapsto \vf_0 \pm 2\pi/3$. The evolution of $\vf_0$ is given by $\vf_0 = p_0 t/3mr^2 + \vf_0(0) + 2 n \pi/3 \;(\text{mod}\; 2\pi)$ where $n = n_2 - n_1$ and $n_{1,2}$ are the `greatest integer winding numbers' of the trajectory around the $\vf_{1,2}$ cycles. Here, $p_0$ is the conserved CM momentum $3 m r^2 \dot \vf_0$.

\subsection{Dynamics on the $\vf_1$-$\vf_2$ torus}
\label{e:s-dyn-on-torus}

Confining ourselves to the motion on the $\vf_1$-$\vf_2$ torus we have the conserved relative energy $E = T + V$ in addition to the conserved energy $3 m r^2 \dot \vf_0^2/2$ of CM motion:
	\beq
	T = \fr{1}{3} m r^2 \left[  \dot \vf_1^2 + \dot \vf_2^2 + \dot \vf_1\dot \vf_2 \right] 
	\quad \text{and} \quad
	V = g \left[3 - \cos \vf_1  - \cos \vf_2  - \cos( \vf_1 +  \vf_2) \right].
	\eeq
Hamilton's equations (with $1 \leftrightarrow 2$ for the other pair)
	\beq
	mr^2\dot \vf_{1} =  2 p_{1} - p_{2} 
	\quad \text{and} \quad
	\dot p_1 = - g \left[ \sin \vf_1 + \sin ( \vf_1 + \vf_2) \right]
	\label{e:3rotors-CM-EOM-phasespace}
	\eeq
follow from the canonical Poisson brackets and
	\beq
	H = \fr{1}{m r^2}(p_1^2 + p_2^2 - p_1 p_2) + V.
	\label{e:hamiltonian-3rotor-phi1phi2space-full}
	\eeq
Extrema of $V(\vf_1,\vf_2)$ lead to six static solutions: (1) the `ferromagnetic ground state' G $((\vf_1, \vf_2) = (0,0))$ where all three rotors coincide ($\tht_1 = \tht_2 = \tht_3$), (2) the three `first excited states' D $((0,\pi), (\pi,0)$ and $(\pi,\pi))$ where two rotors coalesce while the third lies diametrically opposite to them ($\tht_1 = \tht_2 = \tht_3 + \pi$, $\tht_2 = \tht_3 = \tht_1 + \pi$ and $\tht_3 = \tht_1 = \tht_2 + \pi$) and (3) the two `second excited states' T $(\pm 2\pi/3, \pm 2\pi/3)$ where the three bodies lie at vertices of an equilateral triangle ($\tht_1 = \tht_2 + 2\pi/3 = \tht_3 + 4\pi/3$ and $\tht_2 \leftrightarrow \tht_3$). Upon including the CM motion, we get uniformly rotating versions of the above static configurations. G is linearly stable and supports small oscillations with two equal frequencies $\om_0 = \sqrt{3g/mr^2}$. There is one unstable direction around the Ds with growth rate $\om_0$ and one stable eigendirection corresponding to the oscillation frequency $\om_0/\sqrt{3}$. The two triangular `2nd exited states' have two unstable directions, each with growth rate $\om_0/\sqrt{2}$.

As in the planar restricted 3 body problem, the relative motion of the 3 rotor problem (\ref{e:3rotors-CM-EOM-phasespace}) occurs on a 4d phase space but admits only one known constant of motion (\ref{e:hamiltonian-3rotor-phi1phi2space-full}). However, at asymptotically low and high energies, an additional constant of motion is present as in the double pendulum. When $E \gg g$, the  kinetic term far exceeds $V$ and the rotors rotate almost uniformly as $p_{1,2}$ are nearly conserved. At very low energies ($E \ll g$) we have small oscillations around G.

\section{Pendulum and isosceles periodic solutions}
\label{s:pendula-isosceles}

Just as the Euler-Lagrange solutions of the gravitational 3-body problem arise from Keplerian orbits, we seek solutions of the 3 rotor problem that arise from reductions to one degree of freedom. We find two such families: pendula and isosceles breathers.

{\bf Pendulum solutions:} Here, two rotors are assumed to form a `bound pair' with fixed angular separation. Consistency requires their separation to vanish and the equations reduce to the two-rotor pendulum. There are three such families depending on which pair coalesce. Suppose the first two rotors coincide ($\tht_1 = \tht_2$ or $\vf_1 = 0$) so that $\dot \vf_1 \equiv 0$ (or $p_2 - 2 p_1 \equiv 0$). [The other two possibilities are given by $\vf_2 \equiv 0$ and $\vf_1+\vf_2 \equiv 0$]. $\vf_2$ evolves like a pendulum $m r^2 \ddot \vf_2 =  - 3 g \sin \vf_2$ with the conserved energy $E = \ov{3} m r^2 \dot \vf_2^2 + 2g (1 - \cos \vf_2)$ and admits librational ($0 \leq E < 4g$) and rotational ($E > 4g$) orbits with period diverging at $E = 4g$. The pendulum orbits foliate three 2d `pendulum submanifolds' of the 4d phase space defined by relations such as $\vf_1 = 0$ and $p_2 = 2p_1$. Each such periodic pendulum orbit gives rise to a periodic solution of the 3-rotor problem after including the center of mass motion.

{\bf Isosceles breathers:} Here, one rotor is always midway between the other two: $\tht_1 - \tht_2 = \tht_2 - \tht_3$ etc. In this case we obtain a single evolution equation $m r^2 \ddot \vf = - g (\sin \vf + \sin 2\vf)$ for $\vf = \vf_1 = \vf_2$. As with the pendula, every periodic solution of this equation leads to a periodic solution of the 3-rotor problem.
The ground state G is the breather solution with $E = 0$ while D is a breather solution with $E = 4g$. Librational breathers around G have energies $0 \leq E \leq 9g/2$ while those around D exist for $4g \leq E \leq 9g/2$. The time periods of both families grow monotonically and diverge at the $E = 9g/2$ separatrix. Above this energy we only have rotational breathers whose periods decrease with $E$. 

\subsection{Stability of pendula and breathers via monodromy}

By considering small perturbations to a $\tau$-periodic solution, we associate to it a $4 \times 4$ monodromy matrix $M$ whose eigenvalues $\la$ govern its stability.  In fact, the characteristic Lyapunov exponents are $\mu = \log |\la|/\tau$. As with any two-degrees-of-freedom Hamiltonian system, two eigenvalues of $M$ are unity while the other two must be reciprocals \cite{hadjidemetriou}. Since $M$ is real, the reciprocal pair must be of the form $e^{\pm i \phi}$ or $\la^{\pm 1}$ for $\phi, \la \in \mathbb{R}$. Consequently, a pair of Lyapunov exponents vanish with the other two adding up to zero. The orbit is stable if the stability index $\sigma = \tr M - 2$ satisfies $|\sigma| \leq 2$ and unstable if $|\sigma| = |\la + 1/\la| > 2$.

\begin{figure}	
	\centering
	\begin{subfigure}[t]{6cm}
		\centering
		\includegraphics[width=6cm]{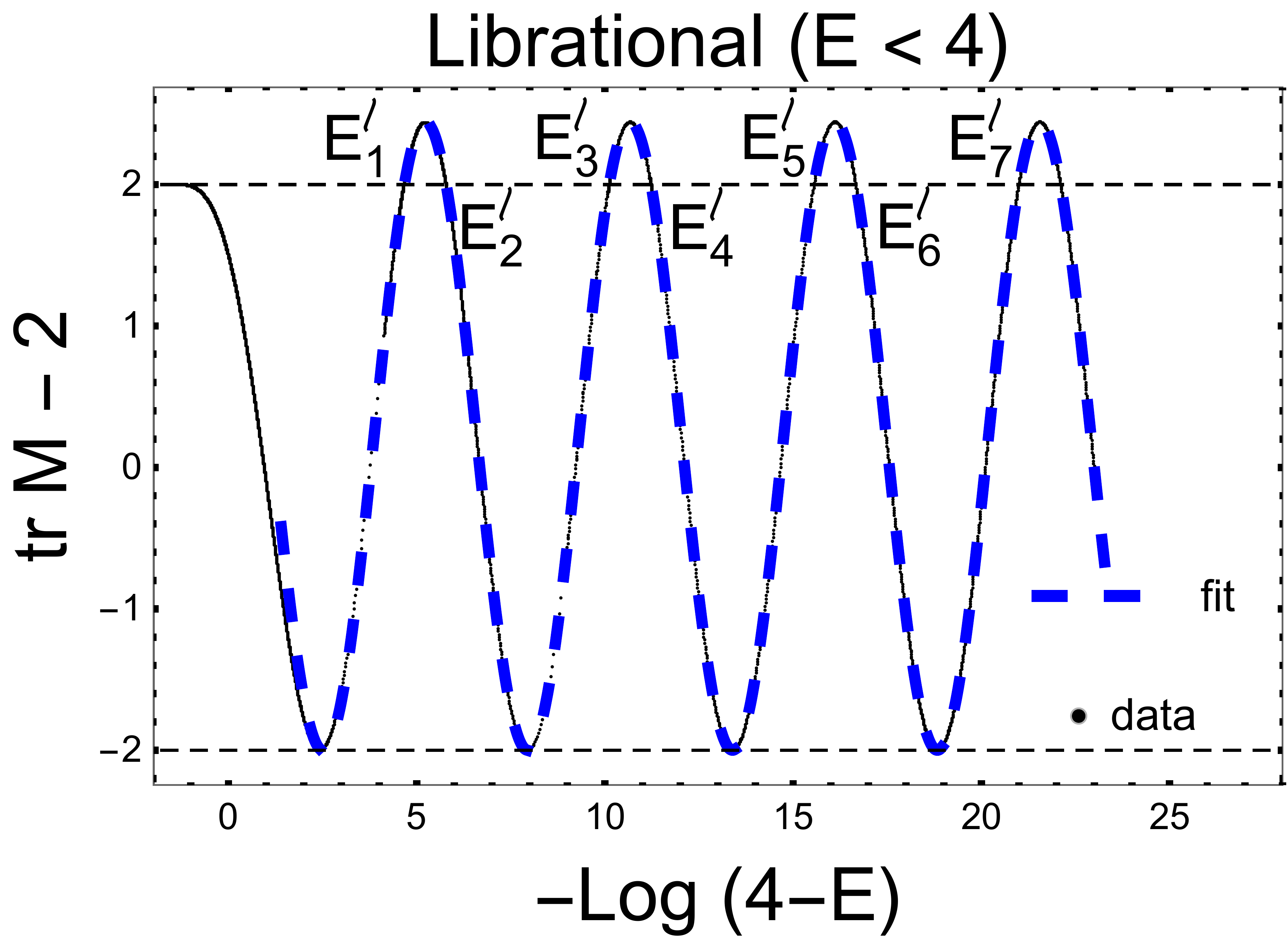}
	\end{subfigure}
\quad
	\begin{subfigure}[t]{6cm}
		\centering
		\includegraphics[width=6cm]{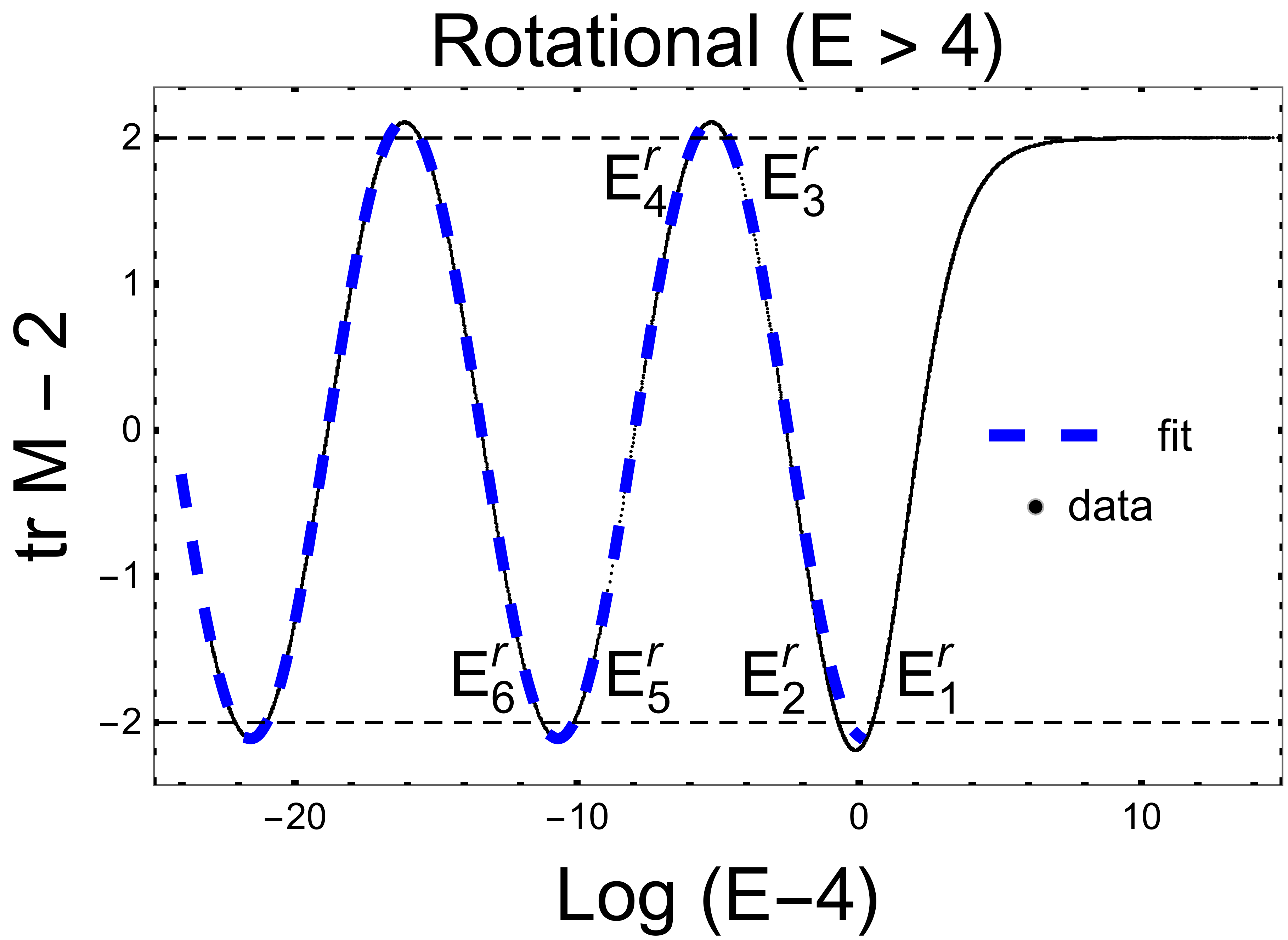}
	\end{subfigure}
	\caption{\footnotesize Numerically obtained stability index $\sigma = \tr M -2$ of periodic pendulum solutions showing asymptotically periodic stable $(|\sig| \leq 2)$ to unstable $(|\sig| > 2)$ transitions on a logarithmic energy scale accumulating at $E = 4$. Eq. (\ref{e:fit-trace-monodromy-lib-rot}) fits the data as $E \to 4^\pm$.}
	\label{f:monodromy-evals}
\end{figure}

{\bf \fl Stability of pendulum solutions:} We evaluate $M$ for the pendula numerically by regarding it as a fundamental matrix solution to the linearized equations. In this case, the neutrally stable 2d eigenspace of $M$ (corresponding to the eigenvalue one) admits a simple interpretation: it is tangent to the pendulum submanifold $(0, \vf_2, p_1, 2p_1)$. Moreover, we find that pendula are stable for low energies $0 \leq E \leq E^\ell_1 \approx 3.99$ and high energies $E \geq E^\err_1 \approx 5.60$ (both in units of $g$) with $M \to I$ as $E \to 0, \infty$. On the other hand, the pendula repeatedly alternate between stable and unstable as the energy approaches four (see Fig.~\ref{f:monodromy-evals}). This  results in an accumulation of stable $\leftrightarrow$ unstable transition energies as $E \to 4^\pm$. In more detail, in the librational regime, the stable phase $[0, E_1^\ell]$ is followed by an unstable one $(E_1^\ell, E_2^\ell)$ which is then followed by another stable phase $[E_2^\ell, E_3^\ell]$ leading to an infinite sequence of successively narrower stable and unstable energy intervals accumulating at $E = 4$. A similar phenomenon is found in the rotational regime when $E \to 4^+$. We will see in \S \ref{s:transition-chaos-global-chaos} that this phenomenon is associated with a rather sharp transition to chaos at $E \approx 4$. Interestingly, the lengths of the stable energy intervals are asymptotically found to be equal on a log scale as are those of the unstable intervals. In fact, we find that the stability index (see Fig. \ref{f:monodromy-evals}) is well approximated  by the following periodic functions of $\log |4-E|$ as $E$ approaches $4$:
	\beq
	\sigma \approx 
	\begin{cases}
	2.22 \cos \left( \fr{2 \log (4 - E)}{\sqrt{3}} + .24 \right) + .22 \quad \text{as} \quad E \to 4^- \quad \text{and} \\
	- 2.11 \cos \; \left( \fr{\log (E - 4)}{\sqrt{3}}  - .12 \right) \quad \text{as} \quad E \to 4^+.
	\end{cases}
	\label{e:fit-trace-monodromy-lib-rot}
	\eeq
A somewhat similar singly-infinite sequence of transitions was found by Yoshida \cite{yoshida84} in the two dimensional anharmonic oscillator $H = p_1^2+p_2^2 + (q_1^4 + q_2^4)/2 + \eps q_1^2 q_2^2$ when the coupling constant $\eps \to \infty$.

{\bf \fl Stability of Isosceles solutions:} The stability of isosceles solutions qualitatively differs from that of pendula: there is just one unstable to stable transition occurring at $E \approx 8.97$. Moreover, we find that both families of librational solutions are unstable with $\sig$ growing monotonically and diverging as $E \to 4.5^-$. Thus, unlike low energy pendula, low energy breathers are unstable despite being small oscillations around the stable ground state G. $\sig$ grows monotonically from $-\infty$ to $2$ for $4.5 \leq E < \infty$ so that rotational breathers go from being unstable to stable at $E \approx 8.97$ where $\sig = -2$. This stability of isosceles solutions is also reflected in the origin of the Poincar\'e section on the $\vf_1 = 0$ surface being a hyperbolic fixed point at low energies and an elliptic fixed point at high energies (see Fig.~ \ref{f:psec}).

\section{Transition to chaos and global chaos}
\label{s:transition-chaos-global-chaos}

\begin{figure}	
	\centering
	\begin{subfigure}[t]{7.5cm}
		\centering
		\includegraphics[width=7.5cm]{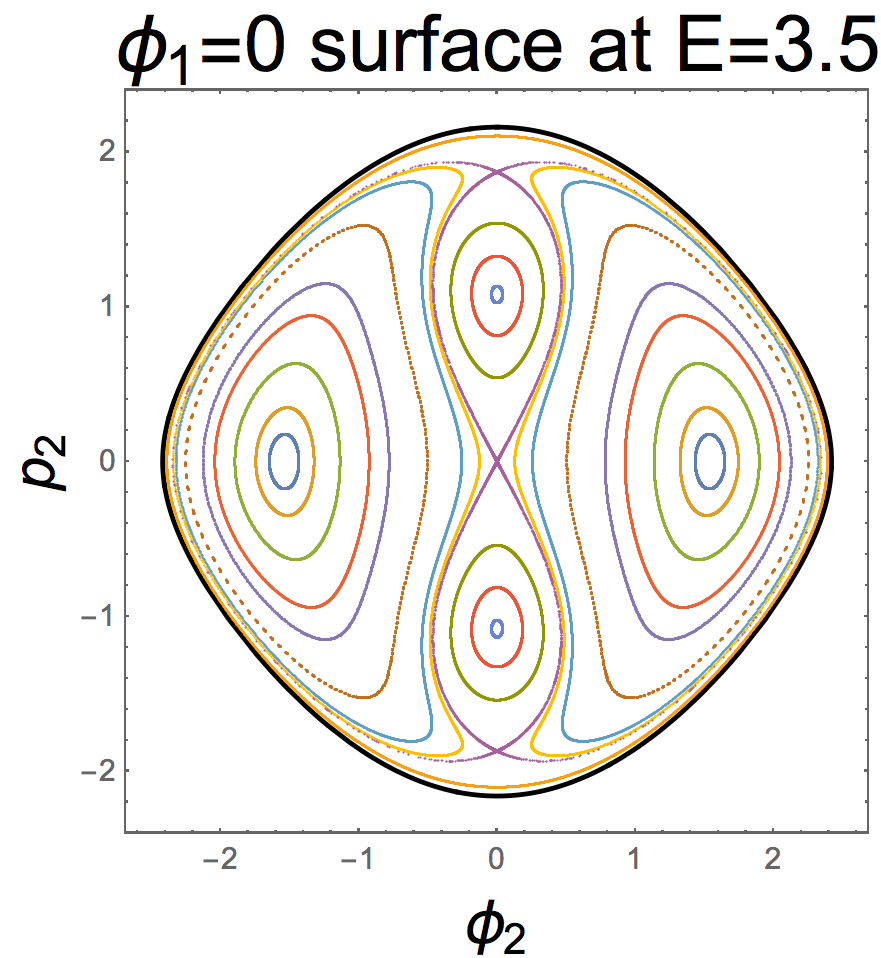}
	\end{subfigure}
\quad
	\begin{subfigure}[t]{7.5cm}
		\centering
		\includegraphics[width=7.5cm]{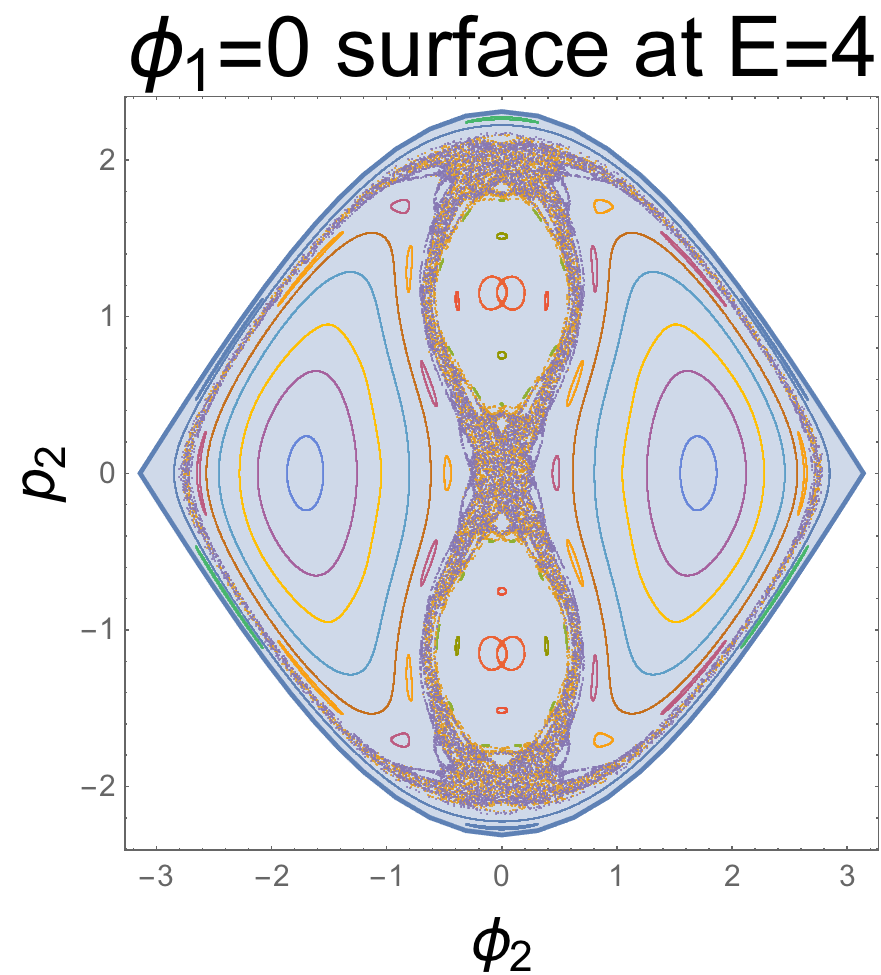}
	\end{subfigure}
\quad
	\begin{subfigure}[t]{7.5cm}
		\centering
		\includegraphics[width=7.5cm]{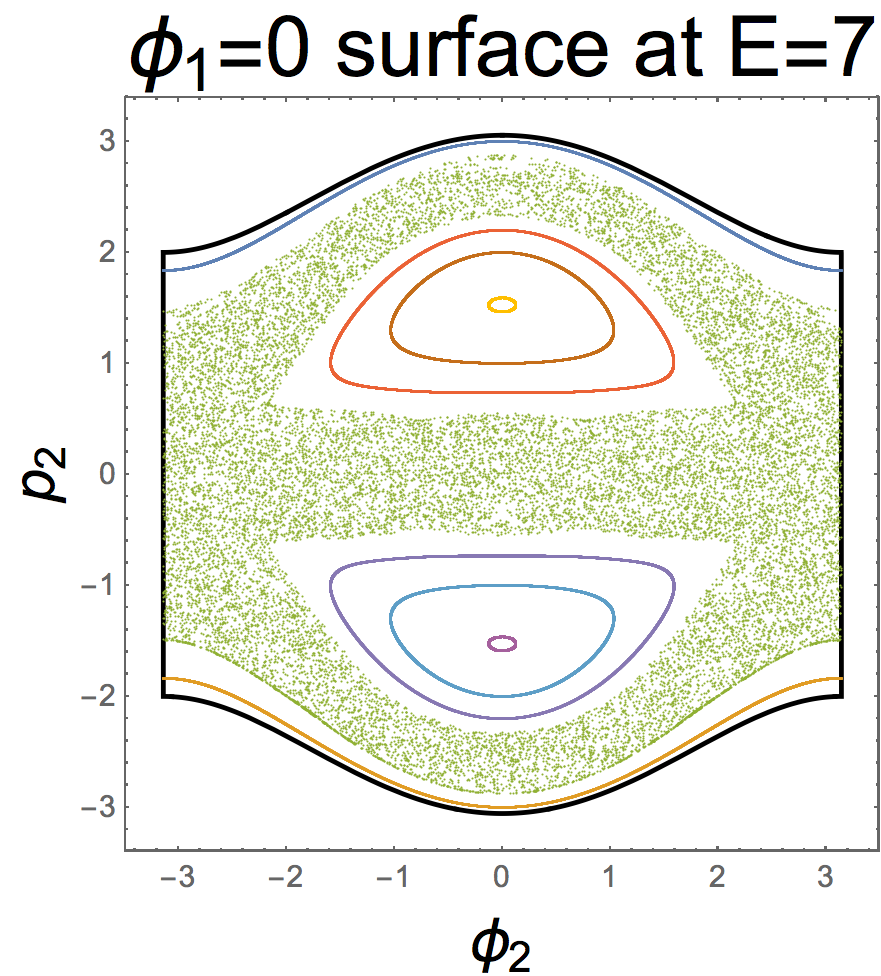}
	\end{subfigure}
\quad
	\begin{subfigure}[t]{7.5cm}
		\centering
		\includegraphics[width=7.5cm]{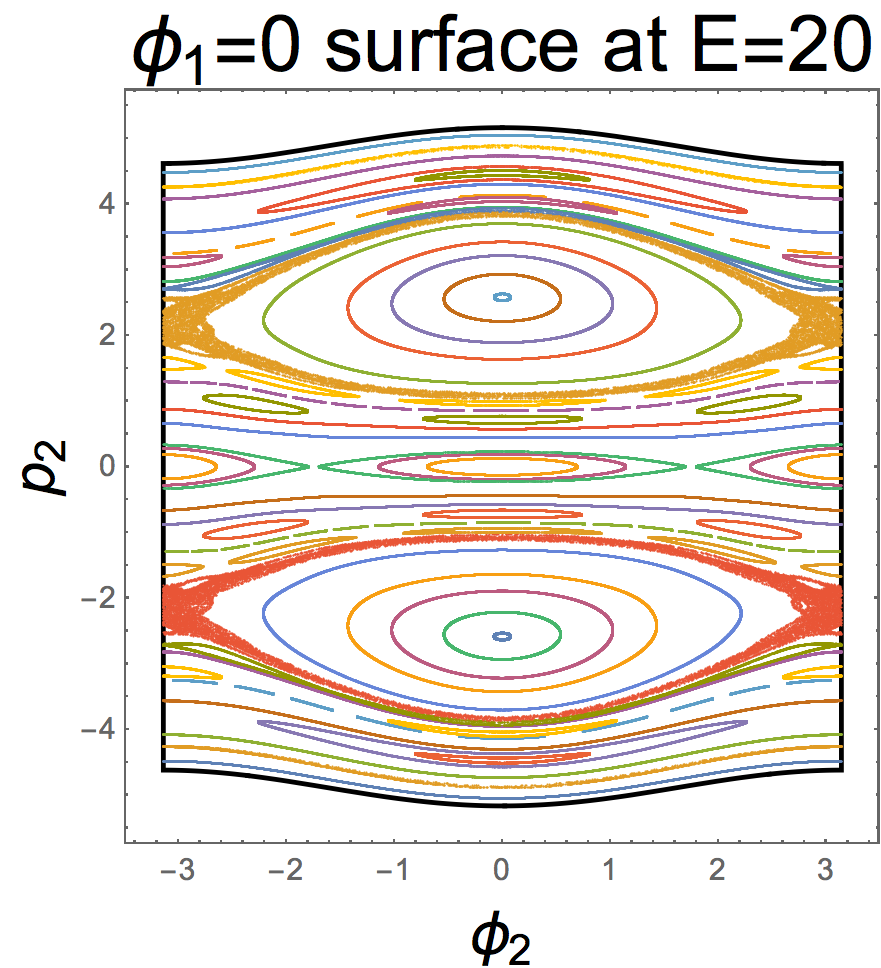}
	\end{subfigure}
	\caption{\footnotesize Poincar\'e sections at various energies on the $\vf_1=0$ surface parameterized by $\vf_2$ and $p_2$. The boundary of the energetically allowed region is a pendulum solution. Distinct sections are colored differently. Breaking of $\vf_2 \to -\vf_2$ and $p_2 \to -p_2$ symmetries is seen at $E = 4$. The fixed point at the origin corresponds to an isosceles periodic solution and goes from hyperbolic to elliptic with increasing energy.}
	\label{f:psec}
\end{figure}

Poincar\'e sections reveal interesting transitions from integrability at asymptotically low and high energies to chaos at intermediate energies in the three rotor problem. We consider 2d Poincar\'e surfaces such as $\vf_1 = 0$ parametrized by $\vf_2$ and $p_2$ with $p_1$ determined by conservation of energy and record a scatter plot of the Poincar\'e return map by numerical solutions of Hamilton's equations. We call a Poincar\'e section `regular' if it is supported on a finite union of points or 1d curves: they arise from periodic and quasi-periodic trajectories. By contrast, a `chaotic' section is one that explores a two-dimensional region. We call the union of all chaotic sections on a Poincar\'e surface at energy $E$ the chaotic region of that Poincar\'e surface.

To obtain Poincar\'e sections, we use explicit and implicit Runge-Kutta schemes as well as symplectic partitioned Runge-Kutta. When there is sensitivity to initial conditions (ICs), different schemes produce trajectories that deviate after some time. Nevertheless, all schemes are found to produce nearly the same Poincar\'e sections for evolution over adequately long times. Moreover, the degree of chaos on various Poincar\'e surfaces such as $\vf_1 = 0$, $\vf_2 = 0$, $p_1 = 0$ and $p_2 = 0$ is found to be qualitatively similar for all ICs considered. 

We find that for $E \lesssim 4$ (in units of $g$), all Poincar\'e sections on the surface `$\vf_1 = 0$' are almost regular and are symmetric under $\vf_2 \to - \vf_2$ and $p_2 \to - p_2$ (see Fig. \ref{f:psec}). Chaotic sections seem to first appear at $E \approx 4$ accompanied by a spontaneous breaking of both these discrete symmetries. While the $\vf_2 \to - \vf_2$ symmetry is restored for $E \gtrsim 4.4$, the $p_2 \to - p_2$ symmetry remains broken. At very high energies, the latter is expected since particles either rotate clockwise or counter-clockwise.

\begin{figure}	
	\centering
	\includegraphics[width=17cm]{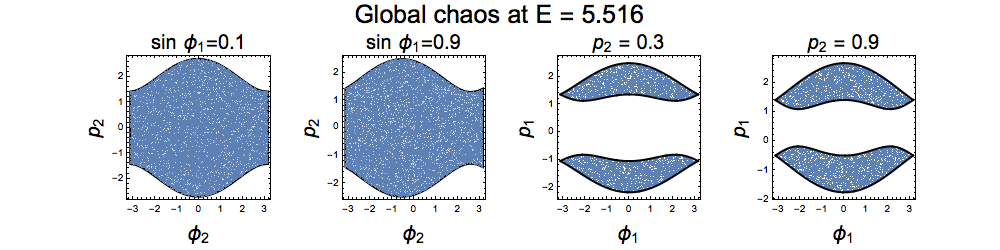}
	\caption{\footnotesize The chaotic region fills up the energetically allowed portion of several Poincar\'e surfaces at $E \approx 5.5$ indicating global chaos.}
	\label{f:global-chaos}
\end{figure}

At intermediate energies $E \gtrsim 4$, we find that chaotic sections from distinct ICs are practically the same while for higher energies, we need to form unions of chaotic sections to find the chaotic region when trajectories are evolved up to $t = 10^5$ in units of $\sqrt{mr^2/g}$. Furthermore, the area of the chaotic region as a fraction of the energetically allowed area of the Poincar\'e surface roughly increases with energy until it saturates (see Fig. \ref{f:psec}). For $5.33 \lesssim E \lesssim 5.6$, the chaotic region fills up all of the energetically allowed part of the Poincar\'e surface. This corresponds to a band of global chaos that is seen on other Poincar\'e surfaces as well (see Fig.~\ref{f:global-chaos}). As the energy is increased further, chaotic sections are supported on progressively thinner bands indicative of the emergence of an additional conserved quantity as $E \to \infty$. 

\begin{figure}	
	\centering
	\includegraphics[width=8cm]{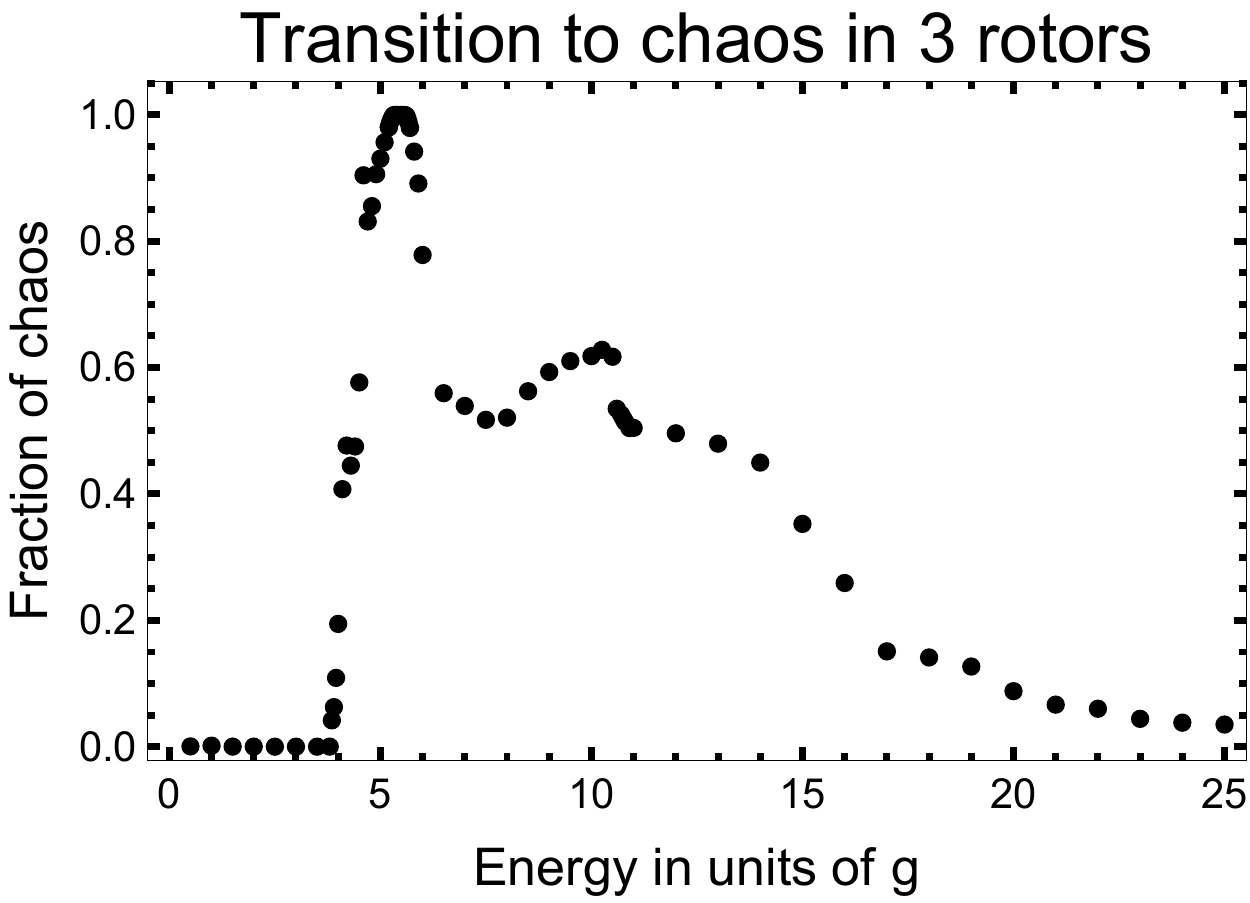}		
	\caption{\footnotesize Fraction $f$ of energetically allowed region on $\vf_1 = 0$ surface occupied by chaotic sections as a function of energy. There is a sharp transition to chaos at $E \approx 4$, a gradual return to regularity as $E \to \infty$ and an indication of global chaos around $E = 5.5$.}
	\label{f:chaos-vs-egy}
\end{figure}

To quantify the foregoing qualitative observations, we exploit the near constancy of the density of points in chaotic sections on the `$\vf_1 = 0$' surface to calculate the `fraction of chaos' $f(E)$ defined as the fraction of the energetically allowed area covered by the chaotic region (see Fig. \ref{f:chaos-vs-egy}). Though chaos does develop even at energies $E \lesssim 4$ (along the periphery of the four lobes in Fig. \ref{f:psec}), the fraction of the area of the Poincar\'e surface occupied by chaotic sections is negligible as indicated by $f \approx 0$ for $E \lesssim 4$. In this restricted sense, we see a rather sharp transition to chaos at $E \approx 4$: $f \sim 10^{-4}$ at $E = 3.8$ while $f \approx .04$, $.06$, $.11$ and $.2$ at $E = 3.85$, $3.9$, $3.95$ and $4$. As seen in Fig.~\ref{f:chaos-vs-egy}, $f$ rises dramatically and reaches $f \approx 1$ in the phase $5.33 \lesssim E \lesssim 5.6$ of global chaos. It then drops gradually to zero with increasing energy. The above sharp transition to chaos is somewhat uncommon  among KAM systems where invariant tori gradually break down. We find that this transition to chaos is also encoded in the Gaussian curvature $R$ of the Jacobi-Maupertuis metric \cite{gskhs-3body}: $R > 0$ in the energetically allowed region for $E < 4$ but takes on either sign when $E > 4$ \cite{gskhs-3rotor}. As already noted in \S \ref{s:pendula-isosceles}, $E = 4$ is also the accumulation point of stable to unstable transitions in both librational and rotational pendulum solutions. Thus, this accumulation of phase transitions in the pendulum orbits appears to be a novel signature of the sharp transition to chaos in the three rotor problem.

In recent times, there has been interest in choreographies in the classical 3 and $n$ body problems. A choreography is a periodic solution where all particles move on the same physical curve equally separated in time, e.g., Lagrange equilateral solutions and the figure 8 in the three body problem \cite{chenciner-montgomery}. Remarkably, we find that the elliptic fixed points at the centers of the left and right lobes in the low energy ($E \lesssim 5.33$) `$\vf_1 = 0$' Poincar\'e surfaces provide ICs leading to choreographies in the 3 rotor problem (see Fig. \ref{f:psec}). This will be elaborated upon in a forthcoming article \cite{gskhs-3rotor}.

{\bf Acknowledgements:} We thank K G Arun, M Berry, A Chenciner, S Dattagupta, S R Jain, A Lakshminarayan, T R Ramadas, M S Santhanam and an anonymous referee for useful comments and references. This work was supported in part by the Infosys Foundation.


\small
\begin{thebibliography}{99}

\bibitem{gutzwiller-three-body} Gutzwiller, M. C., {\it Moon-Earth-Sun: The oldest three-body problem}, Reviews of Modern Physics, {\bf 70}, 589 (1998). 

\bibitem{chenciner-montgomery} Chenciner, A. and Montgomery, R., {\it A remarkable periodic solution of the three-body problem in the case of equal masses}, Ann. of Math. Second Series, {\bf 152}(3), 881-901 (2000).

\bibitem{gskhs-3body} Krishnaswami, G. S. and Senapati, H., {\it Curvature and geodesic instabilities in a geometrical approach to the planar three-body problem}, J. Math. Phys. {\bf 57}, 102901 (2016).

\bibitem{sachdev} Sachdev, S., {\it Quantum phase transitions}, Cambridge University Press, Cambridge (2008).

\bibitem{gskhs-3rotor} Krishnaswami, G. S. and Senapati, H., {\it Classical three rotor problem: periodic solutions, stability and chaos}, arXiv:1811.05807, (2018).

\bibitem{sondhi} Sondhi, S. L., Girvin, S. M., Carini, J. P. and Shahar, D., {\it Continuous quantum phase transitions}, Rev. Mod. Phys., {\bf 69}(1), 315 (1997).

\bibitem{saari-xia} Saari, D. G. and Xia, Z., {\it Off to infinity in finite time}, Notices of the AMS, {\bf 42}, 538 (1993). 

\bibitem{hadjidemetriou} Hadjidemetriou, J. D., {\it Periodic orbits in gravitational systems}, In Chaotic Worlds: From Order to Disorder in Gravitational N-Body Dynamical Systems, Proceedings of the Advanced Study Institute, Steves, B. A., Maciejewski, A. J. and  Hendry, M. (eds.), 43-79 (2006).

\bibitem{yoshida84} Yoshida, H., {\it A Type of Second Order Ordinary Differential Equations with Periodic Coefficients for which the Characteristic Exponents have Exact Expressions}, Celest. Mech. {\bf 32}, 73-86 (1984).




\end{thebibliography}
\end{document}